\documentclass[conference]{IEEEtran}
\IEEEoverridecommandlockouts
\usepackage[T1]{fontenc}
\usepackage[utf8]{inputenc}

\usepackage{xcolor}
\definecolor{MXY}{RGB}{252,8,235}
\usepackage{cite}
\usepackage{amsmath,amssymb,amsfonts}
\usepackage{amsmath}

\usepackage{algorithm}
\usepackage{graphicx}
\usepackage{textcomp}
\usepackage{subcaption} 
\usepackage{algpseudocode} 
\usepackage{hyperref}
\usepackage{xcolor}
\def\BibTeX{{\rm B\kern-.05em{\sc i\kern-.025em b}\kern-.08em
    T\kern-.1667em\lower.7ex\hbox{E}\kern-.125emX}}

\newtheorem{remark}{Remark}

\begin{document}
\title{Optimal RIS Placement in Multi-User MISO Systems with User Randomness}


\author{\IEEEauthorblockN{Abhishek Rajasekaran, Mehdi Karbalayghareh, Xiaoyan Ma, David J. Love, and Christopher G. Brinton}
\IEEEauthorblockA{{Electrical and Computer Engineering, Purdue University, IN 47906, USA} \\
Email:\{\href{mailto:rajasek1@purdue.edu}{rajasek1}, \href{mailto:mkarbala@purdue.edu}{mkarbala}, \href{mailto:ma946@purdue.edu}{ma946}, \href{mailto:djlove@purdue.edu}{djlove},
\href{mailto:cgb@purdue.edu}{cgb}\}@purdue.edu
}}
\maketitle


\begin{abstract}
It is well established that the performance of reconfigurable intelligent surface (RIS)-assisted systems critically depends on the optimal placement of the RIS. Previous works consider either simple coverage maximization or simultaneous optimization of the placement of the RIS along with the beamforming and reflection coefficients, most of which assume that the location of the RIS, base station (BS), and users are known. However, in practice, only the spatial variation of user density and obstacle configuration are likely to be known prior to deployment of the system. Thus, we formulate a non-convex problem that optimizes the position of the RIS over the expected minimum signal-to-interference-plus-noise ratio (SINR) of the system with user randomness, assuming that the system employs joint beamforming after deployment. To solve this problem, we propose a recursive coarse-to-fine methodology that constructs a set of candidate locations for RIS placement based on the obstacle configuration and evaluates them over multiple instantiations from the user distribution. The search is recursively refined within the optimal region identified in each stage to determine the final optimal region for RIS deployment.  Detailed numerical results are presented to corroborate our findings.
\end{abstract}

\section{Introduction}

Reconfigurable intelligent surfaces (RIS) have emerged as a promising solution to enhance the performance of a network\cite{Basar2019IEEEAccess,Wu2020Towards, Liu2021PrincOpportunities}. An RIS embedded with a large number of small reconfigurable elements typically operates as a passive reflector that is connected to a smart controller and is capable of adjusting its reflection coefficients based on channel state information (CSI), if available, so that the desired signals are combined constructively and the interfering signals destructively at the receivers\cite{pathloss2,Wu2021RISToturial}. In many cases, its main role in a network is to mitigate blockages caused by buildings and other obstacles\cite{coverage1,coverage2}, a capability that is particularly valuable for higher-carrier-frequency systems, which are known to suffer from severe blockage issues \cite{power1,6gbrinton}. Given that improving system performance and coverage is the primary utility of an RIS, finding the right place to deploy it in a network is one of the most important and practical problems to address\cite{locbasis,locbasis2}.

\subsection{Related Work}
Recent literature on RIS placement addresses the problem from various perspectives. \cite{imperfectcsiloc} proposes a joint optimization of RIS placement, base station (BS) beamforming, RIS reflection coefficients, and CSI error to maximize the weighted sum rate (WSR) of the system for fixed user locations. In addition, \cite{indoorloc} discusses optimizing placement over the coverage area for the special case of indoor communications at THz frequencies. Furthermore, \cite{coverageloc} postulates placement optimization over the probability of coverage for random users, BSs, and obstacles, but produces mostly theoretical results and does not engage in performance maximization. \cite{loc3,loc4,loc6} also consider the placement problem for specific applications such as high-speed trains, aerial backhaul systems, and unmanned aerial vehicle (UAV) communications. 

The performance of an RIS-assisted network is enhanced by jointly optimizing the BS beamforming and the reflection coefficients at the RIS (called joint beamforming). However, since a single beamformer and phase configuration cannot effectively serve all possible user combinations from a distribution, this optimization is mostly performed after the system is deployed.  \cite{weightedsumrate} presents one of the most accepted joint beamforming algorithms that maximizes WSR. Some other strategies are listed in \cite{beam2,beam3,jung}. The machine learning based approaches are not considered because of their dependence on vast training data, which we might not be able to provide in practical scenarios. 

Despite extensive existing research on RIS-based systems, no existing work optimizes RIS placement by maximizing coverage, while also accounting for performance, under user randomness.

\subsection{Overview of Methodology and Contributions}
To address this practical yet unexplored problem, we propose to maximize the expected minimum signal-to-interference-plus-noise ratio (SINR) of the system with respect to the user distribution, under the assumption that joint beamforming is performed after deployment of RIS at its optimal location. The max-min objective inculcates fairness by extending the coverage of the BS to most of the network. However, it is impossible to solve this optimization directly, since the computation of the objective requires knowledge of the beamformer and reflection coefficients, which in turn can only be practically determined after deployment. 

Hence, we construct a discrete set of candidate RIS locations for each instantiation of a set of users from the user distribution. For each of these candidates, joint beamforming is performed to compute the optimal beamformer and phases. Using these, for each instantiation of users, the candidate location that maximizes the minimum SINR of the system is chosen to be a feasible solution. Then, this search is further refined within the optimal region identified in each recursive call to determine the final optimal region for RIS deployment.

\begin{figure}[httb]
\centerline{\includegraphics[width=0.45\textwidth]{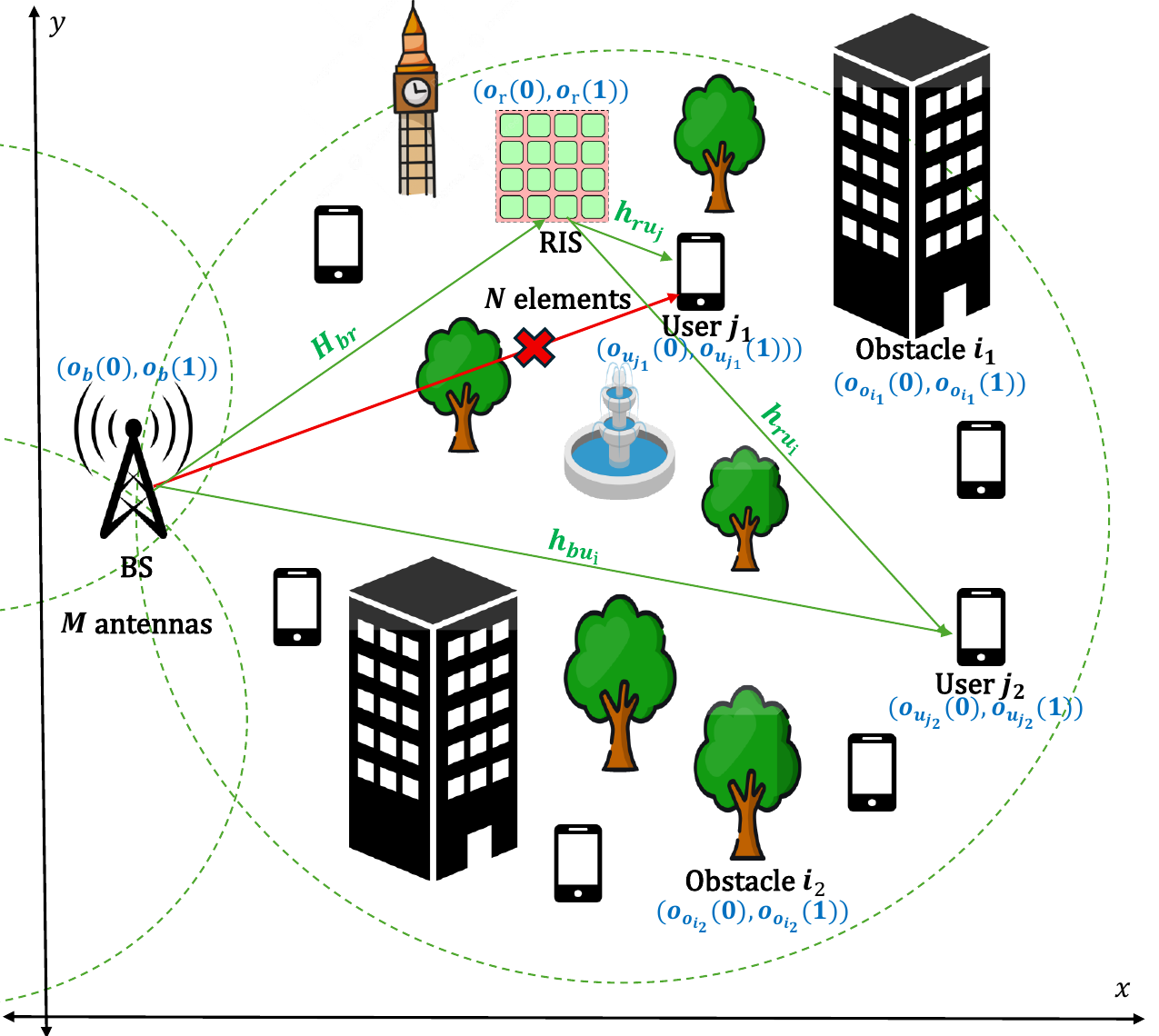}}
\caption{RIS-assisted MU-MISO system with obstacles. \label{fig:sysmodel}}
\end{figure} 

The proposed technique is the first work, to the best of our knowledge, that simultaneously achieves multiple typical practical requirements of an RIS-assisted system, such as maximizing performance by exploiting the spatial variation of user density, enhancing coverage by exploiting the obstacle configuration, and computing the optimal RIS location without the deterministic user locations. The superiority of the proposed scheme was validated using simulations.

\section{System Model}
\label{sec:system-model}

Consider a multi-antenna BS at $\mathbf{o_b}\in\mathbb{R}^2$ serving multiple cells (assumed circular for simplicity) as illustrated in Fig.~\ref{fig:sysmodel}. We assume that all cells operate using orthogonal frequencies, which makes it sufficient to investigate just one cell.  

In the cell under investigation, a multi-user multiple-input single-output (MU-MISO) system is considered where the BS with $M$ antennas serves multiple single-antenna users. We also assume that $I$ obstacles hinder signal delivery from the BS. For modeling purposes, these obstacles are represented by two main types: (1) circular obstacles such as pillars and (2) wall-type obstacles. Let $\{\mathbf{o_o}\} \in \mathbb{R}^2$ denote the locations of the centers of all obstacles. Circular-type obstacles are associated with another set to denote their radii $\{r_{ci}\}$. In contrast, wall-type obstacles are associated with their lengths $\{l_{wi}\}$ and orientations $\{\theta_{wi}\}$. Furthermore, we assume that all obstacles are identical about the z-axis as shown in Fig.~\ref{fig:sysmodel} (which makes it sufficient to investigate just with its x, y coordinates). Thus, $(\{\mathbf{o_o}\},\{r_{ci}\},\{l_{wi}\}, \{\theta_{wi}\})$ together is defined as the ``obstacle configuration.'' From the practical standpoint, this information is fully available to us once the area of deployment (like a specific school) is decided (then the positions of the buildings, pillars, etc. in the school are known). Thus, in this work, we assume that the obstacle configuration is fully known.

On the other hand, practically, we would not know the exact locations of users ahead of deployment. We might only know the spatial variability in user density across the cell. This is mathematically formulated as the user distribution.  Unlike conventional spatial distributions like the 2D Gaussian or uniform models, the Poisson Point Process (PPP) provides greater versatility in representing the spatial variability of users in dense environments \cite{coverageloc}. Hence, the positions of the users are represented by a homogeneous PPP $\Psi_{u} = \{\mathbf{o_u}\} \in \mathbb{R}^2$ with density $\lambda_u$. Let $K$ denote the number of users in the system corresponding to a specific instance of this PPP.

The direct link from BS to user $k \in \{1, \ldots, K\}$ is modeled using the Rayleigh fading model as
\begin{equation}
   \mathbf{h}_{b{u}_k} = \sqrt{\beta_{bu_k}}\tilde{\mathbf{h}}_{b{u}_k} ,
\label{eq:1}
\end{equation}
where $\beta_{bu_k}$ and $\tilde{\mathbf{h}}_{b{u}} \in \mathbb{C}^{M\times 1}$ denote the path loss and the unit-variance circular symmetric complex Gaussian (CSCG) component of the direct link, respectively. 

Consider an RIS with $N$ reflecting elements located at $\mathbf{o_r} \in \mathbb{R}^2$. The links between the BS and RIS and between RIS and the $k^{th}$ user are modeled using the Rician fading model, which are respectively given as 
\begin{align}
    \mathbf{H}_{br} &= \sqrt{\beta_{br}}\left(\sqrt{\frac{T_1}{1+T_1}}\mathbf{\hat{H}}_{br}+ \sqrt{\frac{1}{1+T_1}}\tilde{\mathbf{H}}_{b{r}}\right),
\end{align}
\begin{align}
    \mathbf{h}_{ru_k} = \sqrt{\beta_{ru_k}}\left(\sqrt{\frac{T_2}{1+T_2}}\mathbf{\hat{h}}_{ru_k}+ \sqrt{\frac{1}{1+T_2}}\mathbf{\tilde{h}}_{ru_k}\right) ,
\end{align}
where $\beta_{br}$ and $\beta_{ru_k}$ denote the path loss components of the BS-RIS and RIS-User~$k$ links, respectively. $\tilde{\mathbf{H}}_{b{r}} \in \mathbb{C}^{N\times M}$ and $ \mathbf{\tilde{h}}_{ru_k}\in \mathbb{C}^{N\times1}$ denote the unit-variance CSCG components of the corresponding links, and $\mathbf{\hat{H}}_{br}=\mathbf{a}_N(\vartheta
)\mathbf{a}_{M}^H(\psi)$ and $\mathbf{\hat{h}}_{ru_k} = \mathbf{a}_N(\zeta_k)$ denote the deterministic components of the links. Here, $\mathbf{a}_i$ is the steering vector of size $i$ for $i\in\{N,M\}$ and $\vartheta,  \psi,\zeta_k$ are the angular parameters. $T_1$ and $T_2$ are the Rician parameters that govern the balance between the steering vector components and the CSCG components. The reflection matrix of the RIS is given by
\begin{align}
    \boldsymbol{\Theta} &=  \text{diag}(\boldsymbol{\theta}),\\
    \text{where  } \boldsymbol{\theta} &= [\theta_{1},\ldots,\theta_{N}]^T,
\end{align}
in which, $\theta_n = e^{j\varphi_n}$ is the phase-shift introduced at the $n^{th}$ element of RIS, for all $1 \le n \le N$. Let the combined signal transmitted from the BS be $\mathbf{x} = \sum_{k=1}^{K} \mathbf{w}_k s_k$, where $\mathbf{w}_k \in \mathbb{C}^{M \times 1}$ is the transmit beamformer and $s_k$ the unit-power transmit symbol corresponding to the $k^{th}$ user. The signal received at user $k$ can be expressed as

\begin{align}
y_k &= \mathbf{h}_{{bu}_k}^{\mathrm{H}} \mathbf{x} + \mathbf{h}_{ru_k}^{\mathrm{H}} \boldsymbol{\Theta} \mathbf{H}_{br} \mathbf{x} + n_k \nonumber \\
&= \left( \mathbf{h}_{{bu}_k}^{\mathrm{H}} + \mathbf{h}_{ru_k}^{\mathrm{H}} \boldsymbol{\Theta} \mathbf{H}_{br} \right) \sum_{k=1}^{K} \mathbf{w}_k s_k + n_k,
\end{align}
where \( n_k \sim \mathcal{CN}(0, \sigma^2) \) is the AWGN at the $k^{th}$ user. For computational convenience, the indirect links are redefined using \( \mathbf{H}_{r,k} = \text{diag}(\mathbf{h}_{{ru}_k}^{\mathrm{H}}) \mathbf{H}_{br} \). Thus, the revised expression for the received signal at user~$k$ is given by 

\begin{equation}
y_k = \left( \mathbf{h}_{{bu}_k}^{\mathrm{H}} + \boldsymbol{\theta}^{\mathrm{H}} \mathbf{H}_{r,k} \right) \sum_{k=1}^{K} \mathbf{w}_k s_k + n_k,
\end{equation}
Let $\mathbf{h}_k \triangleq (\mathbf{h}_{{bu}_k}^H+\boldsymbol{\theta}^H \mathbf{H}_{r,k})$ denote the end-to-end channel for user~$k$. Thus, the SINR at user~$k$ is
\begin{equation}
\gamma_k = \frac{\left|  \mathbf{h}_k \mathbf{w}_k \right|^2}{\sum_{i \ne k} \left| \mathbf{h}_k \mathbf{w}_i \right|^2 + \sigma^2}.
\label{eq:SINR}
\end{equation}

\section{Problem Formulation}
\label{sec:problem-formulation}
Clearly, for a system as shown in Fig.~\ref{fig:sysmodel}, an RIS is required to extend the coverage of the BS to most of the cell. This must be done, also ensuring good performance. This problem is formulated as
\begin{align*}
    \mathbf{P1.1}:\quad\max_{\mathbf{o_r}}&\quad \mathbb{E}_{\mathbf{o_u}}\left[\min \gamma_k \right]\nonumber\\
    =\max_{\mathbf{o}_r}&\quad \mathbb{E}_{\mathbf{o}_u}\left[\min \frac{\left|  \mathbf{h}_k \mathbf{w}_k \right|^2}{\sum_{i \ne k} \left| \mathbf{h}_k \mathbf{w}_i \right|^2 + \sigma^2}\right] \nonumber
\end{align*}
in which we assume that $\boldsymbol{\theta}$ in $\mathbf{h}_k$ and the accumulated beamformer matrix $\mathbf{W} = [\mathbf{w_1},\cdots,\mathbf{w_K}] \in \mathbb{C}^{M\times K}$ are optimized using
\vspace{-2ex}
\begin{subequations}
\begin{align}
    \mathbf{P1.2}:\quad (\boldsymbol{\theta},\mathbf{W}) = \arg \max_{\boldsymbol{\theta,\mathbf{W}}} &\sum_{k=1}^{K}\frac{1}{K}\log(1+\gamma_k)\nonumber\\
        s.t. \quad &\sum_{k=1}^{K}||\mathbf{w_k}||_2^2 \le P_{max},
\label{eq:inneropti1}\\
        &|\theta_n| = 1, \forall n = 1,\cdots, N.
\label{eq:inneropti2}
\end{align}
\end{subequations}
The objective function in $\mathbf{P1.2}$ is the average rate or WSR with equal priority assigned to all users.  Eq.~\eqref{eq:inneropti1} enforces the total transmit power constraint $P_{max}$ for the system. Eq.~\eqref{eq:inneropti2}, on the other hand, defines the feasible set for the RIS reflection coefficients by enforcing unit amplitude at all $N$ elements.

In problem $\mathbf{P1.1}$, we intend to find that RIS placement that maximizes the expected SINR of the most-affected user among the $K$ users (that has the least SINR) to enhance the coverage of the system. This is called `max-min' solution which is the most ``fair'' one. This problem of finding the most fair solution for the placement of RIS is denoted as the ``outer optimization''. It should be noted that this `expectation' of SINR is taken over the user distribution. We perform this outer optimization assuming that joint beamforming is performed after the RIS is deployed at its optimal location. This part of the problem, $\mathbf{P1.2}$, is denoted as the ``inner optimization''. 

Many researchers consider this multilayered structure of the formulation to be the `ideal' one that is closest to the practical scenario. However, because of the difficulty in directly solving it, existing works like \cite{indoorloc,coverageloc,imperfectcsiloc} simplify the formulation in various ways. In the next section, we propose an elegant technique to directly tackle the practical hierarchical formulation.  


\section{Proposed Solution}
\label{sec:proposedsol}
In order to solve $\mathbf{P1.1}$, we need the optimal beamformer and phases from $\mathbf{P1.2}$. Therefore, in this section, we first consider solving the joint beamforming problem $\mathbf{P1.2}$ and then the outer optimization problem $\mathbf{P1.1}$.

\subsection{The Inner Optimization Problem}
\label{subsec:inneropti}

\subsubsection{Transformed Problem} Using the Lagrangian dual transform \cite{fp}, $\mathbf{P1.2}$ is reduced to a sum-of-ratios problem by introducing a set of auxiliary variables $\boldsymbol{\alpha} = \{\alpha_1,\dots,\alpha_K\}$

\begin{subequations}
\begin{align}
\mathbf{P2.1}:\quad\max_{\boldsymbol{\alpha},\mathbf{W},\boldsymbol{\theta}}&  \sum_{k=1}^K\Big[ \frac{1}{K} (\log(1+\alpha_k) - \alpha_k + \frac{(1+\alpha_k)\gamma_k}{1+\gamma_k}\Big]\nonumber \\
s.t. \quad &\sum_{k=1}^{K}\|\mathbf{w}_k\|_2^2 \leq P_{max},\\
&|\theta_n|=1, \quad \forall n = 1,\cdots, N,\\
&\alpha_k\ge0, \quad \forall k = 1,\cdots, K.
\end{align}
\end{subequations}

Clearly, the problem is equivalent to $\mathbf{P1.2}$ as long as $\alpha_k = \gamma_k$. Then, the quadratic transform \cite{qt} is applied to decouple the ratios (SINR) in $\mathbf{P2.1}$ by introducing another set of auxiliary variables $\boldsymbol{\beta} = \{\beta_1,\dots,\beta_K\}$
\begin{subequations}
\begin{align}
\mathbf{P2.2}:\quad\max_{\boldsymbol{\alpha},\boldsymbol{\beta},\mathbf{W},\boldsymbol{\theta}}&  \sum_{k=1}^K\Big[ \frac{1}{K} (\log(1+\alpha_k) - \alpha_k) \nonumber\\
+ & 2\sqrt{\frac{1}{K} (1+\alpha_k)} \mathcal{R}e\Big\{ \beta_k^*\mathbf{h}_k\mathbf{w}_k \Big\} \nonumber\\
- &|\beta_k|^2 \Big( \sum_{i=1}^K |\mathbf{h}_k\mathbf{w}_i|^2 + \sigma^2 \Big)\Big]\nonumber \\
s.t. \quad &\sum_{k=1}^{K}\|\mathbf{w}_k\|_2^2 \leq P_{max},\\
&|\theta_n|=1, \forall n = 1,\cdots, N,\\
&\alpha_k\ge0, \quad \forall k = 1,\cdots, K.
\end{align}
\end{subequations}

\subsubsection{Iterative Optimization}

In order to solve $\mathbf{P2.2}$, we adopt an alternative optimization (AO) strategy. The auxiliary variables are updated with their optimal values \cite{beamxi} 
\begin{align}
\alpha_k &= \gamma_k,\label{eq:a}\\
\beta_k &= \frac{\sqrt{\frac{1}{K} (1+\alpha_k)}\,\mathbf{h}_k \mathbf{w}_k}{\sum_i |\mathbf{h}_k \mathbf{w}_i|^2 + \sigma^2}.
\label{eq:b}
\end{align}
Looking at the well-studied optimal transmit beamforming problem with constant phases, one can derive the update rule for $\mathbf{W}$ to be the WMMSE solution which is given by
\begin{align}
\mathbf{w}_k &=  \sqrt{(1 + \alpha_k)} \beta_k \Big(\lambda^{\text{opt}} \mathbf{I}_M +\sum_{i=1}^K |\beta_i|^2 \mathbf{h}_i \mathbf{h}_i^H \Big)^{-1} \mathbf{h}_k,\label{eq:W}\\
\lambda^{\text{opt}} &= \min \left\{ \lambda^{\text{opt}} \geq 0 \mid \sum_{i=1}^K |\mathbf{w}_k|^2 \le P_{max} \right\},
\label{eq:lambda}
\end{align}
where $\lambda^{\text{opt}}$ is obtained by solving Eq.~\eqref{eq:lambda} using a bisection search over a finite interval that starts at 0 (upto $10^{9}$, let's say). Finally, one can see that the phase dependent terms in the objective from $\mathbf{P2.2}$ can be shown \cite{weightedsumrate} to be of the form 

\begin{align}
\mathbf{P2.3} \quad    &\min_{\boldsymbol{\varphi}} f_3(\boldsymbol{\varphi}) = (e^{j\boldsymbol{\varphi}})^H\boldsymbol{U}e^{j\boldsymbol{\varphi}} - 2 \mathcal{R}e\{\boldsymbol{v}^He^{j\boldsymbol{\varphi}}\}\\
    \text{where }\quad \boldsymbol{U} &= \sum_{k=1}^K|\beta_k|^2 \sum_{i=1}^K \mathbf{H}_{r,k} \mathbf{w}_i (\mathbf{H}_{r,k} \mathbf{w}_i )^H,\\
    \boldsymbol{v} &= \sum_{k=1}^K\Big(\sqrt{\frac{1}{K}(1+\alpha_k)}\beta^*_k(\mathbf{H}_{r,k}\mathbf{w}_i)\nonumber\\ 
    &- |\beta_k|^2\sum_{i=1}^K (\mathbf{h}^H_{bu,k} \mathbf{w}_i)^*\mathbf{H}_{r,k}\mathbf{w}_i\Big),
\end{align}

Hence, to optimize phases with constant transmit beamforming, we adopt the gradient projection algorithm 
\begin{equation}
\boldsymbol{\varphi} \leftarrow \left(\boldsymbol{\varphi} - \rho \nabla f_3(\boldsymbol{\varphi})\right)^+,
\label{eq:phase}
\end{equation}
where $(.)^+$ denotes the projection of the enclosed set of phases into the region $|\theta_n| =1, \forall n=1,\cdots,N$. Also, $\rho$ is dynamically updated in each iteration for faster convergence.

We alternatively optimize $(\boldsymbol{\alpha},\boldsymbol{\beta},\mathbf{W},\boldsymbol{\theta})$ using Eq. ~\eqref{eq:a},~\eqref{eq:b},~\eqref{eq:W} and~\eqref{eq:phase} iteratively to solve the inner optimization problem.

\subsection{The Outer Optimization Problem}
\label{subsec:outeropti}

Since closed-form solutions for $\mathbf{w}$ and $\varphi$ as functions of the channel in $\mathbf{P1.2}$ is clearly intractable, while solving $\mathbf{P1.1}$, we are instead required to approximate the expectation of SINR over multiple realizations of the user distribution. 

A discrete set of candidate locations for RIS (of cardinality $T$) can be formed using Algorithm~\ref{algo:candidateset} assuming that Circle $C$ to be the entire cell. The main idea behind the algorithm stems from the necessity for the RIS to have a clear Line of Sight (LoS) from the BS to be able to forward the message to all users. In addition, the RIS is placed at least $d_{FF} = \frac{2D^2}{\lambda_c}$ far from the BS (Fraunhofer distance of the BS) \cite{farfield}, so that the RIS does not interfere in the near field of the BS. Here, $\lambda_c$ denotes the wavelength corresponding to the frequency of operation and $D$ denotes the overall aperture size of the BS.
\begin{algorithm}[ht!]
\caption{Proposed Construction of the Candidate set $\mathcal{Q}$}\label{algo:candidateset}
\textbf{Inputs:}
\begin{enumerate}
    \item[(i)] BS location: $\mathbf{o_b}$,
    \item[(ii)] Obstacle configuration: $( \{\mathbf{o_o}\},\{r_{ci}\},\{l_{wi}\}, \{\theta_{wi}\})$,
    \item[(iii)] Circle $C$.
\end{enumerate}
\textbf{Output:} $\mathcal{Q}$
\begin{algorithmic}[1]
    \State Set $\mathcal{Q} = \varnothing$ and $i = 1$.
    \While{$i \leq T$} 
        \State Pick a random $\mathbf{q}_i$ in $C$
        \If{$\mathbf{q_i}$ is at least $d_{FF}$ away from $\mathbf{o_b}$ \textbf{ and }Clear LoS from $\mathbf{o_b}$ to $\mathbf{q}_i$}
            \State Update $\mathcal{Q} = \mathcal{Q} \cup \{\mathbf{q}_i\}$
            \State Set $i = i + 1$
        \EndIf
    \EndWhile
\end{algorithmic}
\end{algorithm}

Furthermore, a set of feasible `max-min' solutions (set $\mathcal{S}$) can be constructed using these candidate sets as outlined in Algorithm~\ref{algo:solset}. For a large number of instantiations of sets of users from the homogeneous PPP distribution, we first form corresponding candidate sets using Algorithm~\ref{algo:candidateset}. Then, joint beamforming is performed assuming that the RIS is placed at each candidate location in each of the candidate sets (one after another). Finally, from each candidate set, that RIS that maximizes the performance of the worst performing user in that set of $K$ users, is added as a solution to $\mathcal{S}$.

\begin{algorithm}[ht!]
\caption{Proposed Construction of the Solution set $\mathcal{S}$}\label{algo:solset}
\textbf{Inputs:}
\begin{enumerate}
    \item[(i)] User distribution,
    \item[(ii)] Circle $C$.
\end{enumerate}
\textbf{Output:} $\mathcal{S}$
\begin{algorithmic}[1]
\State Create multiple instantiations of users from its distribution and use Algorithm \ref{algo:candidateset} to get corresponding candidate sets 
\State For each candidate, in each candidate set, corresponding to an instantiation of users, perform joint beamforming
\State For each instantiation of a set of users, a feasible solution is obtained using $\mathbf{P1.1}$, forming a solution set $\mathcal{S}$ 
                
\end{algorithmic}
\end{algorithm}

\begin{algorithm}[ht!]
\caption{Computing the final solution}\label{algo:globalsol}
\textbf{Inputs:}
\begin{enumerate}
    \item[(i)] Solution set: $\mathcal{S}$,
    \item[(ii)] Initial step size: $d_{start}$,
    \item[(iii)] Precision threshold: $d_p$.
\end{enumerate}
\textbf{Output:}  $\mathbf{\hat{x}_r} = \Call{RecursiveAlgo}{\mathcal{S}, d_{start}}$
\begin{algorithmic}[1]
    \Procedure{RecursiveAlgo}{$\mathcal{S}_{temp}, d$}
        \If{$d < d_p$}
            \State \textbf{return} mode of set $\mathcal{C}_{temp}$
        \Else
            \State Define $\mathcal{C}_{temp} = \{\text{round}(c/d) \cdot d \mid c \in \mathcal{S}_{temp}\}$
            \State Construct $\tilde{\mathcal{S}}$ using Algorithm \ref{algo:solset} for a circle $C$ centered at the mode of $\mathcal{C}_{temp}$ and fixed radius $r$ 
            \State \textbf{return} \Call{RecursiveAlgo}{$\tilde{\mathcal{S}}, d/2$}
        \EndIf
    \EndProcedure
\end{algorithmic}
\end{algorithm}

\subsection{Recursive Clustering}
\label{subsec:globalsol}
It is important to observe that this solution set is not closed under linear combinations, i.e. $s_1\in \mathcal{S}$ and $s_2\in \mathcal{S} \nRightarrow (a.s_1+b.s_2) \in \mathcal{S}, \forall a,b\in \mathbb{R}$. This rules out the usual tendency to compute the mean of all optimal locations in $\mathcal{S}$ to find the `final solution' (mean of solutions not necessarily a solution itself). 
\begin{table*}[t]
    \caption{Time complexity comparison with benchmarks}
    \centering
    \setlength{\tabcolsep}{1.5pt} 
    \begin{tabular}{l c c c}
        \hline
        \textbf{Algorithm} & \textbf{Complexity of $\mathbf{W}$ update} & \textbf{Complexity of $\boldsymbol{\theta}$ update} & \textbf{Overall Complexity Order}\\ 
        \hline
        Proposed & $\mathcal{O}(KM^2)$ & $\mathcal{O}(KN^2\min{\{M,K\}})$ & $\mathcal{O}(I_{\text{iter}}(KM^2 + KN^2\min{\{M,K\}}))$\\
        \cite{imperfectcsiloc} & $\mathcal{O}(K(2M^3 + NM^2))$ & $\mathcal{O}(IG( NM^2 + MN + M))$ & $\mathcal{O}(I_{\text{iter}}(M^3+ MN^2+ K(2M^3 + NM^2) + IG( NM^2 + MN + M) + KMN)$\\
        \cite{indoorloc} & $O(K^2V)$ & $O(K^2JW)$ & $\mathcal{O}(I_{\text{iter}}(K^2V+K^2JW))$\\
        \hline
        \end{tabular}
    \label{tab:complexity_comp}
\end{table*}
Therefore, we propose Algorithm~\ref{algo:globalsol} which is a recursive method to compute the final solution. `Clusters' are formed among the solutions in $\mathcal{S}$ by creating quantized versions of them with a suitable step size, $d_{start}$. Then the cluster with the highest number of entries is declared to be the `optimal cluster'. Then Algorithm~\ref{algo:solset} is repeated for a new candidate set with candidates only within the optimal cluster. This process of looking deeper into the optimal cluster is repeated up to the required precision $d_p$. Hence, the globally optimal region to deploy the RIS is obtained as the square $[\mathbf{\hat{x}_r}(1)- d_p/2,\mathbf{\hat{x}_r}(1)+ d_p/2)\times[\mathbf{\hat{x}_r}(2)- d_p/2,\mathbf{\hat{x}_r}(2)+ d_p/2)$. 


\begin{remark}
The core part of the proposed algorithm only requires the obstacle configuration, channel parameters, and the user distribution. Owing to the separable structure of the problem formulation in $\mathbf{P1.1}$ and $\mathbf{P1.2}$, the algorithm is capable of using any proposed or future joint beamforming technique. And there are many existing joint beamforming algorithms that work under imperfect CSI conditions including \cite{beam3}. In our work, we have used our adaptation of a well-accepted beamforming algorithm \cite{weightedsumrate} that works with perfect CSI. That clearly does not limit the proposed algorithm in any way to work only under perfect CSI conditions. In other words, the proposed approach remains \emph{effective} for any CSI acquisition and RIS training technique with the corresponding channel estimation errors, which jointly affect the SINR expression in Eq.~\eqref{eq:SINR}, prior to its use in our algorithm.
\end{remark}

\subsection{Time Complexity and Efficient Implementation}
The overall computational efficiency of the proposed algorithm is strongly dependent on the efficiency of the inner optimization. The time complexity of updating $\boldsymbol{\alpha}$ and $\boldsymbol{\beta}$ using Eq.~\eqref{eq:a} and~\eqref{eq:b} can be easily shown to be $\mathcal{O}(KMN)$. 

On the other hand, optimization of transmit beamforming from Eq.~\eqref{eq:W} involves matrix inversion, making this step $\mathcal{O}(KMN + I_{\lambda}M^3)$ where $I_{\lambda}$ is the number of iterations for the bisection search for $\lambda$. This step could potentially become a bottleneck for the efficiency of the proposed solution. Hence, we adopt a linear update \cite{weightedsumrate}
\begin{align}
    &\mathbf{w}_k = \frac{1}{L-\lambda}(L\hat{\mathbf{w}}_k-\mathbf{g}_k),\\
    &\lambda = \frac{L}{2}-\frac{1}{2P_{max}}\sum_{k=1}^K||L\hat{\mathbf{w}}_k-\mathbf{g}_k||, 
\end{align}
where $\hat{\mathbf{w}}_k=\mathbf{w}_k+\epsilon(\mathbf{w}_k-\mathbf{w}^{\text{prev}}_k)$ for some $\epsilon>0$. Here, $\mathbf{g}_k= - \frac{\partial f_2}{\partial \mathbf{w}_k}$ and $L$ is set to the Lipschitz constant of $\mathbf{g}_k$. This linear update results in the same WMMSE solution in Eq.~\eqref{eq:W}, but reduces time complexity to a reasonable $\mathcal{O}(KM^2)$.

Similarly, the phase optimization is dominated by the calculation of $\boldsymbol{U}$ and the gradient projection algorithm. The resulting time complexity of the direct implementation is $\mathcal{O}(K^2N^2+I_{\rho}N^2) = \mathcal{O}(K^2N^2)$ where $I_{\rho}$ is the number of iterations spent in calculating the dynamic step size $\rho$. Instead of implementing $\boldsymbol{U}$ directly like \cite{weightedsumrate} or \cite{imperfectcsiloc}, one can also take the summation $\sum_{i=1}^K\mathbf{w}_i\mathbf{w}^H_i$ outside and compute it just once to get a time complexity of $\mathcal{O}(KMN^2)$. Since $K$ is a Poisson random variable in our implementation, we adopt a dynamic strategy to use the direct implementation of $\boldsymbol{U}$ if it is the usual case of $M \ge K$ and the second if $M<K$. This results in the best complexity of $\mathcal{O}(KN^2\min\{K,M\})$.

Hence, the overall complexity of the proposed solution is $\mathcal{O}(I_{RL}\times I_{AO}\times (KMN + KM^2+KN^2\min\{K,M\}))$ where $I_{RL}$ is the number of recursive levels traced in the outer optimization and $I_{AO}$ is the number of inner optimization iterations. It is also important to note that this is a one time cost spent before the RIS assisted system is actually implemented. 

A comparison of the computational efficiency of our method with similar benchmarks is provided in Table \ref{tab:complexity_comp}. Let $I_{\text{iter}}$ uniformly denote the total number of iterations we execute the considered algorithm. For our algorithm, $I_{\text{iter}} = I_{RL}\times I_{AO}$. From Table~\ref{tab:complexity_comp}, \cite{imperfectcsiloc} uses the matrix inversion implementation for transmit beamforming. Also, it assumes that the number of users and their actual locations are precisely known before placement which is not practical. Our implementation is more efficient and clearly more practical. On the other hand, the time complexity of the analog beamforming in \cite{indoorloc} is comparable to that of the beamformer update in our proposed method. The time complexity of $\mathcal{O}(K^2JW)$ for phase optimization is very similar to the complexity of direct implementation, $\mathcal{O}(K^2N^2)$. Ours is a more generalized version of the same order.
\begin{table}[t]
    \caption{Optimal RIS Placement in different scenarios}
    \centering
    \setlength{\tabcolsep}{1pt}  
    \begin{tabular}{l c c c c c}
        \hline
        \textbf{Scenarios} & $\mathbf{\lambda_u}$ & \textbf{Average WSR} & \multicolumn{2}{c}{\textbf{Coverage (\%)}} \\
        & & \textbf{(bps/Hz)} & \textbf{With RIS} & \textbf{Without RIS}\\
        \hline
         1 & 0.009 & 6.7915 &99.68\% & 65.76\%\\
        2 & 0.007 & 7.6556 & 97.91\% & 66.27\%\\
        3 & 0.009 & 6.7522 & 97.21\% & 75.96\%\\
        4 & $\begin{cases}
        0.4, & \text{Circle } C_1\\
        0.04, & \text{Circle } C_2\\
        0.05, & \text{Circle } C_3\\
        0.035, & \text{Circle } C_4\\
        0,&\text{elsewhere}
    \end{cases}$ & 6.6982& 99.69\% & 70.17\%\\
        \hline
    \end{tabular}
    \label{tab:sim_params}
\end{table}

\begin{figure*}
    \begin{subfigure}[b]{0.49\columnwidth}
        \includegraphics[width=\linewidth]{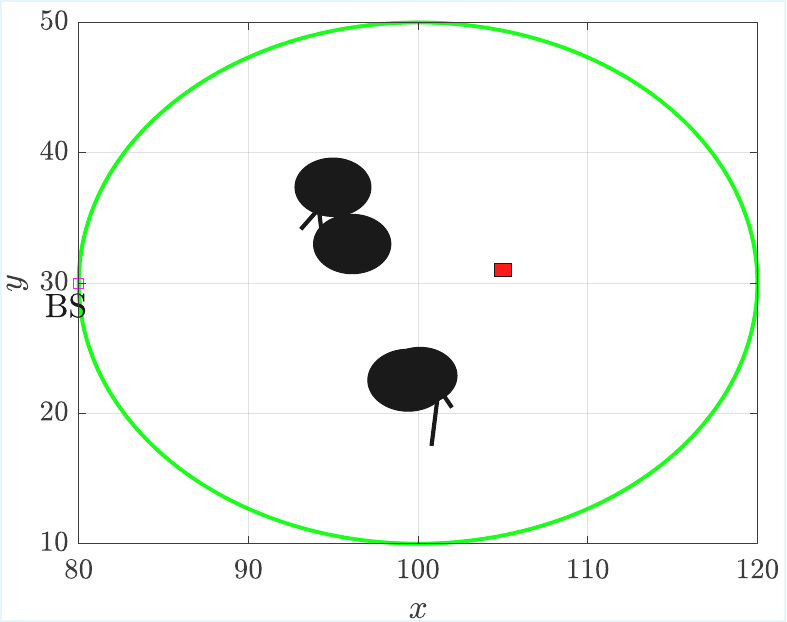}
        \caption{Scenario 1: Final Solution.}
        \label{fig:scenario1finalsol}
    \end{subfigure}
    \hfill
    \begin{subfigure}[b]{0.49\columnwidth}
        \includegraphics[width=\linewidth]{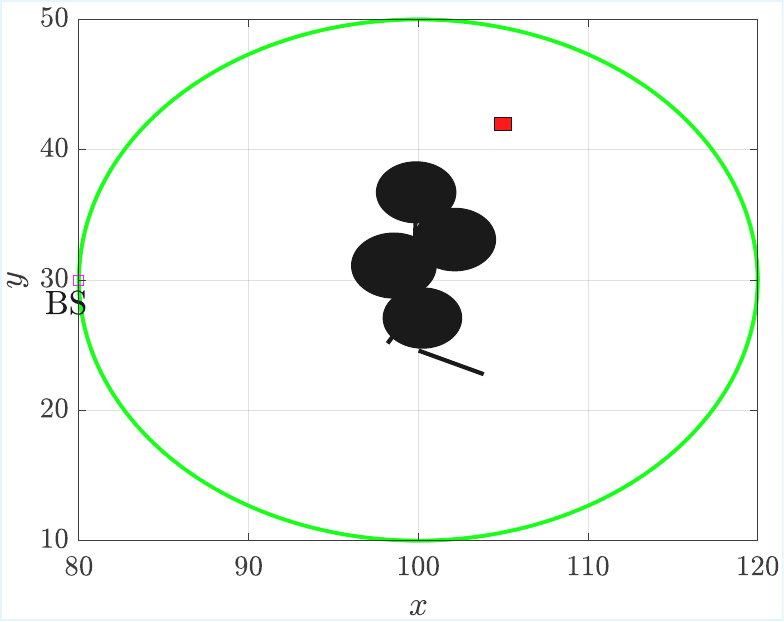}
        \caption{Scenario 2: Final Solution.}
        \label{fig:scenario2}
        \end{subfigure}
\hfill
    \begin{subfigure}[b]{0.49\columnwidth}
        \includegraphics[width=\linewidth]{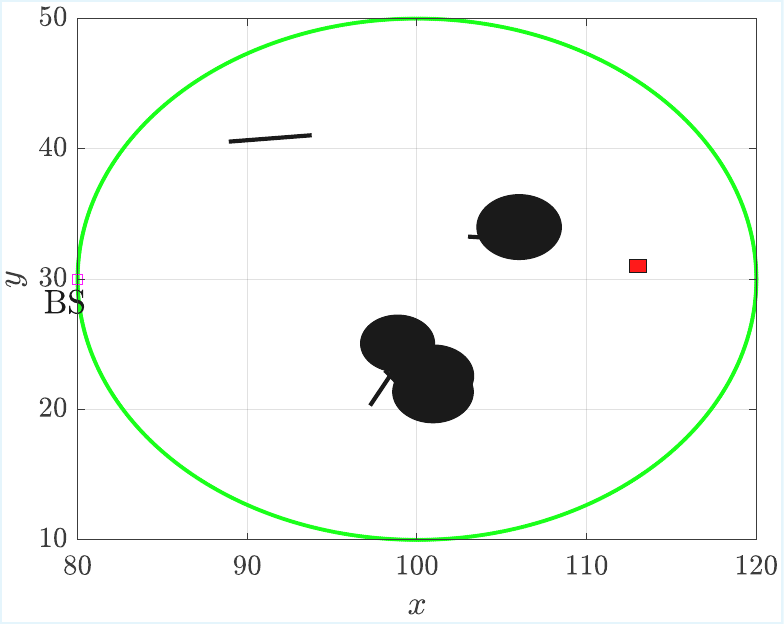}
        \caption{Scenario 3: Final Solution.}
        \label{fig:scenario3}
    \end{subfigure}
    \hfill 
    \begin{subfigure}[b]{0.49\columnwidth}
        \includegraphics[width=\linewidth]{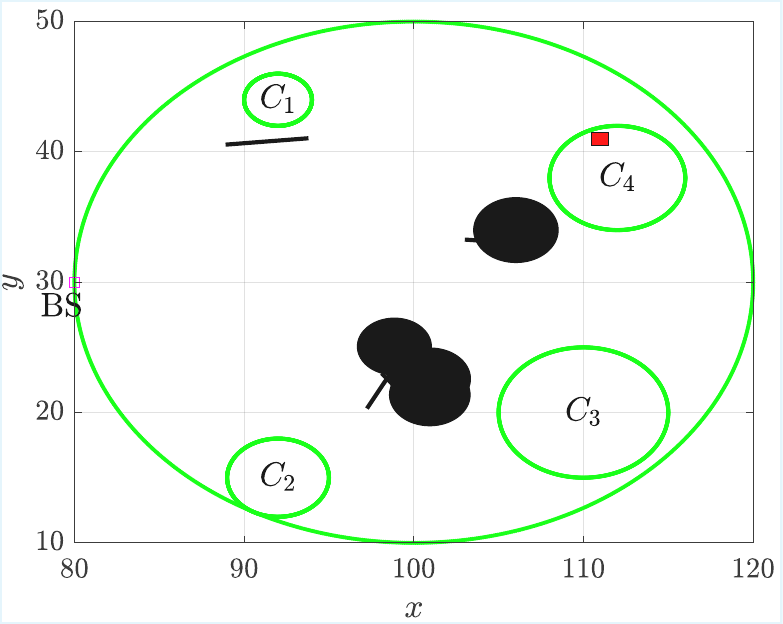}
        \caption{Scenario 4: Final Solution.}
        \label{fig:scenario4}
    \end{subfigure}
     \caption{Optimal RIS placement in different scenarios.}
     \label{fig:allscenarios}
\end{figure*}

\begin{figure*}[t!]
    \centering
    \begin{subfigure}[b]{0.65\columnwidth}
        \includegraphics[width=\linewidth]{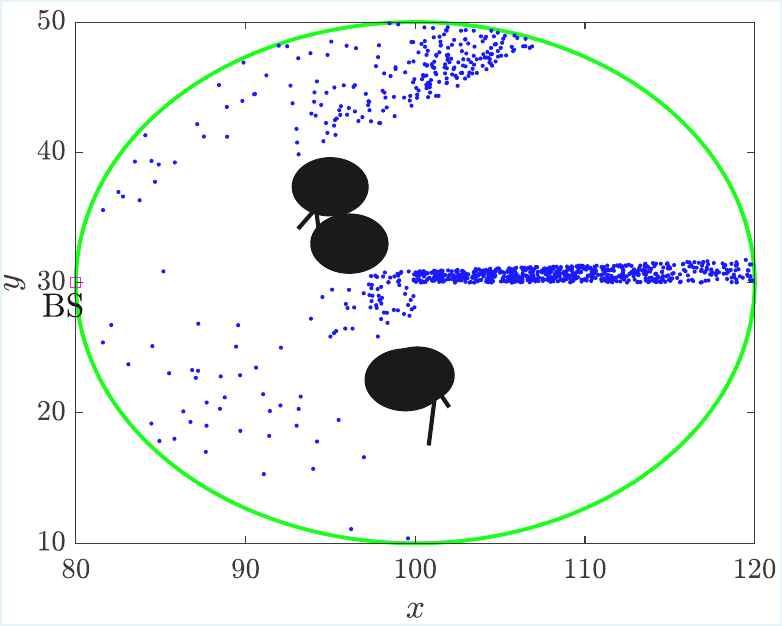}
        \caption{Set of min-max solutions obtained in Recursion Level 1.}
        \label{fig:scenario1scatterplot}
    \end{subfigure}
    \hfill 
     \begin{subfigure}[b]{0.65\columnwidth}
        \includegraphics[width=\linewidth]{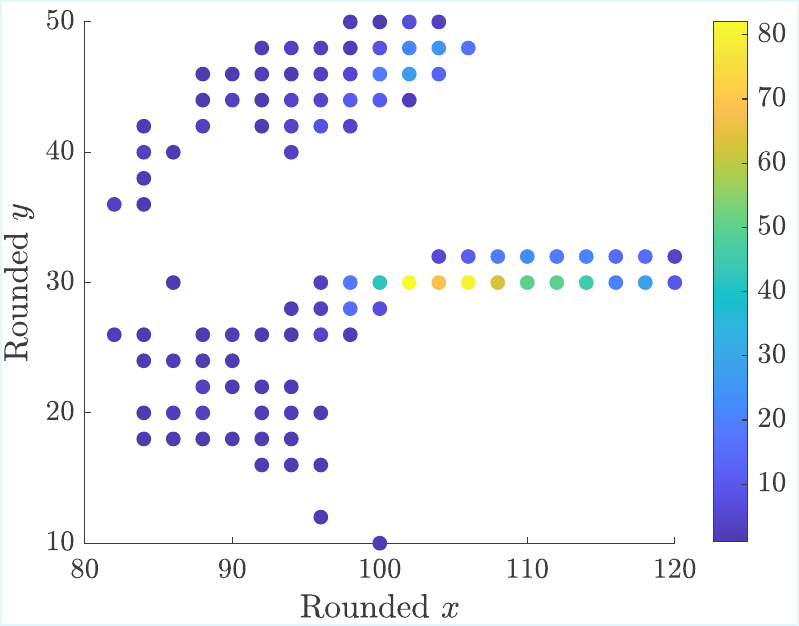}
        \caption{Frequency map of the quantized solution set in Recursion Level 1.}
        \label{fig:scenario1freq1}
    \end{subfigure}
    \hfill
    \begin{subfigure}[b]{0.65\columnwidth}
        \includegraphics[width=\linewidth]{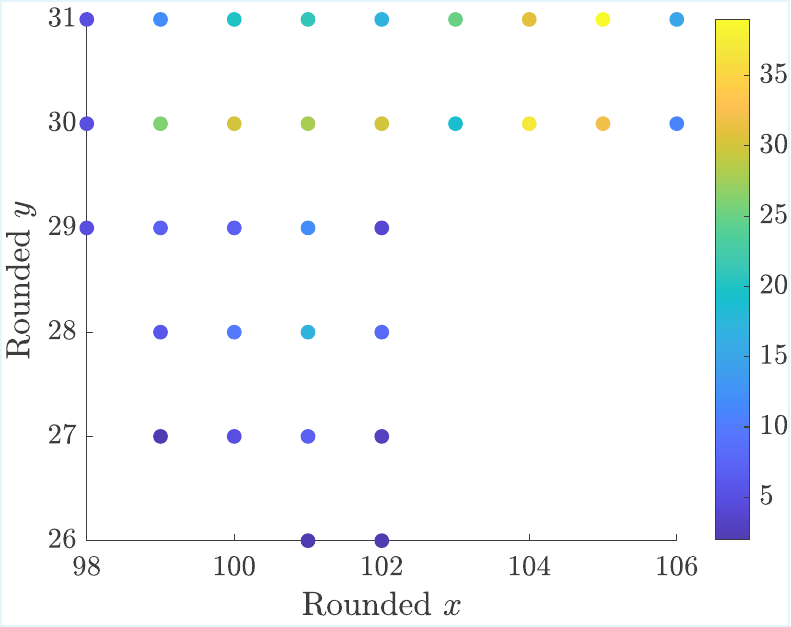}
        \caption{Frequency map of the quantized solution set in Recursion Level 2.}
        \label{fig:scenario1freq2}
    \end{subfigure}
    \caption{Recursive Clustering for Optimal RIS placement in Scenario 1.}
\end{figure*}

\begin{remark}
Since $K$ in our system is random, all analytical results are derived in general ($K$ may be greater than, equal to, or less than the number of transmit antennas $M$). However, it is important to note that when the system is overloaded ($M<K$), the derived WMMSE result is no longer optimal unless it is combined with an appropriate scheduling algorithm, which will select $S \leq M$ users that jointly employs beamforming and hence do not interfere with each other. For scheduling purposes, these $S$ users are treated as a single ``virtual'' user occupying one time slot. The remaining $K-S$ users may interfere with each other and with this virtual user. So, their transmission is handled via orthogonal multiple access (to mitigate the interference). Although for simplicity, we assume that $M\ge K$ in the simulations presented, we emphasize that the proposed algorithm is general and can be effectively integrated with any scheduling algorithm in literature. A detailed investigation is beyond the scope of this paper and is omitted for brevity and space limits.

\end{remark}

\section{Simulation Results}
\label{sec:results}

The simulation parameters used are laid out below, followed by a detailed review of the results obtained. 
\subsection{Simulation Setup}
The path loss parameters $\beta_{br}$, $\beta_{ru_k}$, $\beta_{bu_k}$ are assumed to obey the 3GPP standard for Urban Micro cell (Table B.1.2.1-1 in \cite{3gpp}),
\begin{align}   
    &\beta_{i} = 22\log_{10}(d_i) + 28 + 20\log_{10}(f_c) \text{, }i\in\{br,ru_k\},\\
    &\beta_{bu_k} = 36.7\log_{10}(d_{bu_k}) + 22.7 + 26\log_{10}(f_c),
\end{align}
where $d_i$ represents the Euclidean distance traversed by each link in meters (m) and $f_c$ is the carrier frequency in GHz. $f_c$ is set as $2.4$ GHz. For example, we get a loss of $\beta_{br} = \beta_{ru_k} = 55.6$ dB and $\beta_{bu_k} = 69.3$ dB for $d_{br} = d_{ru_k} = d_{bu_k} =  10$~m. Then the Rician factors are set to $T_1 = T_2 = 10$ dB. The transmit power is set to $P_{max} = 0$ dB and $\sigma^2 = -81$ dBm \cite{pathloss2} unless mentioned otherwise. We also assume that when a link hits an obstacle, the link power is reduced to a negligible level due to the severe penetration loss.


The BS is assumed to be located at $\mathbf{o_b} = (80$ m$,30$ m$)$ at a corner of the cell as in Fig. \ref{fig:sysmodel} and the cell radius is fixed as $R = 20$~m. Furthermore, other system parameters are set to $M = 16, N = 100, d_p = 1$~m for all simulations. Finally, coverage is computed over a large grid of points in the cell (but not on top of any obstacle) with a resolution of $0.1$~m.

\subsection{Optimal RIS Placement}
\label{subsec:results}

Four different scenarios are considered for simulation as shown in Table \ref{tab:sim_params} and Fig.~\ref{fig:allscenarios}. The first three scenarios assume various positions, sizes, and orientations for obstacles that hinder access to a set of users following a homogeneous PPP with a fixed density $\lambda_u$ throughout the circular cell. Finally, a more practical scenario with four user hotspots is considered.

Fig.~3~(a)-(c) present a top-level view of the complete procedure described in Section~\ref{sec:proposedsol} for Scenario~1. Fig.~3~(a) shows a scatterplot of the solution set in recursion level (RL)~1, in which the black objects are the obstacles, the green circle represents the user cluster (the whole cell in this case), and the blue dots represent the solutions. Fig.~3~(b) represents the frequency heatmap of a quantized version of this solution set with $d_{start}=2$~m. To achieve $d_p = 1$~m, another recursive pass of the algorithm is run as shown in Fig.~3~(c). The cluster with the highest RL 2 frequency is declared as the optimal region for RIS deployment (represented by a $d_p \times d_p$ square) in Fig.~2~(a). As per Table~\ref{tab:sim_params}, deploying RIS at this location extends coverage from 65.76\% to 99.68\%  of the cell under consideration. The average WSR performance of deploying RIS at the center of this cluster is also captured in Table \ref{tab:sim_params}. Note that the expected number of users corresponding to this scenario is a practical $E[K] = \lambda_u.\text{Area} = 11.3097$.

\begin{figure}[t!]
\centering
     \begin{subfigure}[b]{0.66\columnwidth}
        \includegraphics[width=\linewidth]{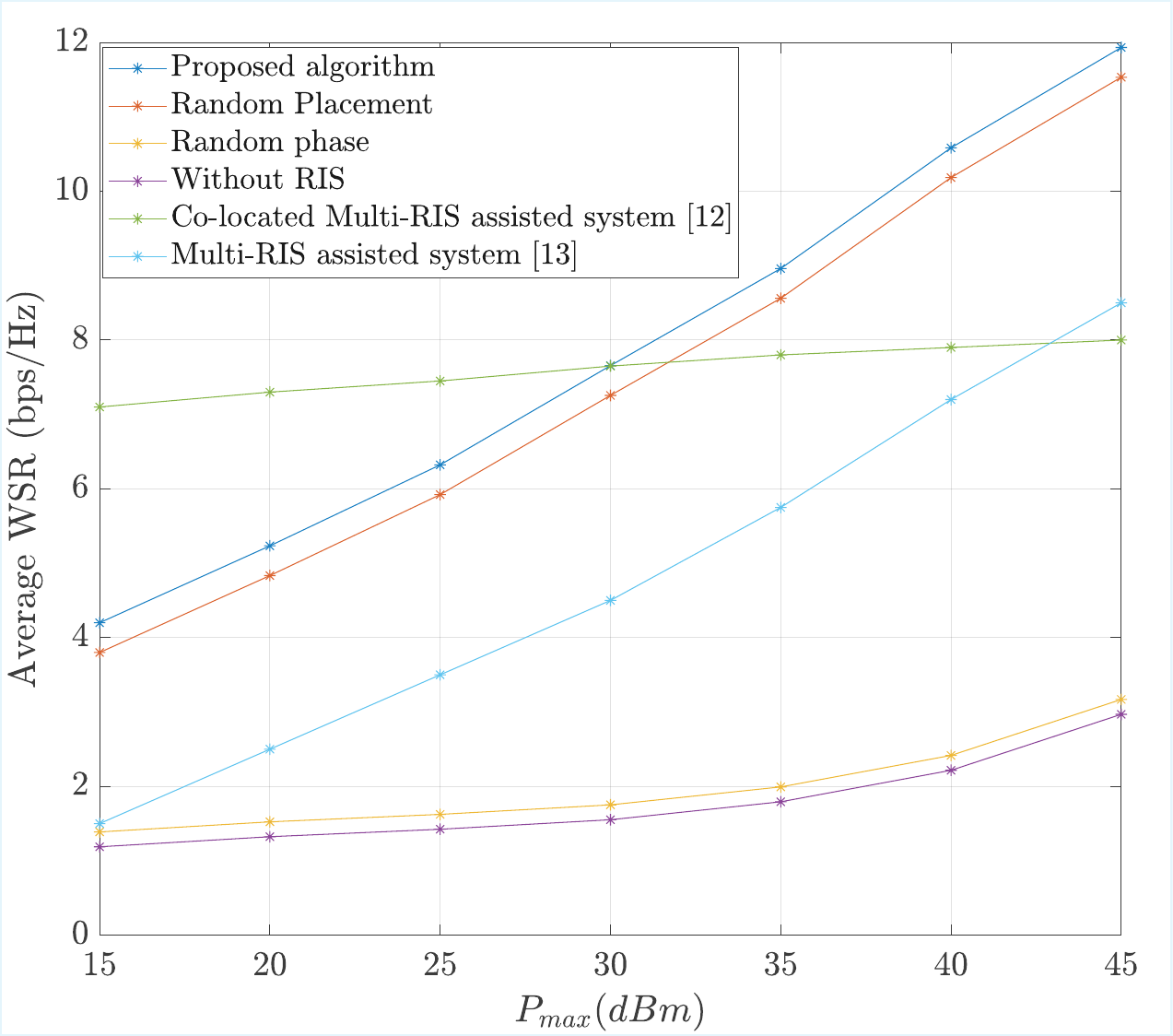}
        \caption{$P_{max}$ characteristic.}
        \label{fig:scenario1freq1}
    \end{subfigure}
  
    \begin{subfigure}[b]{0.70\columnwidth}
        \includegraphics[width=\linewidth]{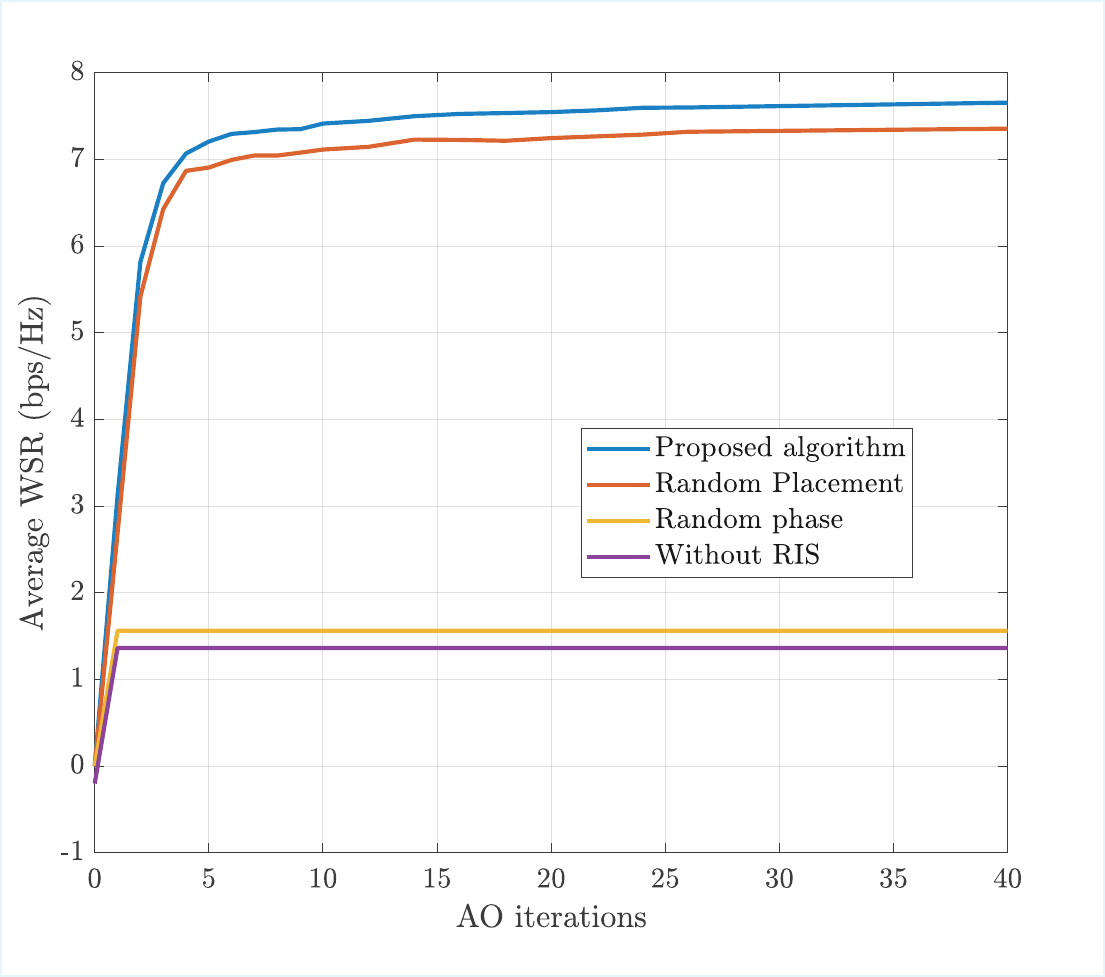}
        \caption{Convergence analysis.}
        \label{fig:scenario4}
    \end{subfigure}  
    \caption{Conventional Performance analysis of proposed algorithm in Scenario 2.}
  \end{figure}  
  
  \begin{figure}
    \begin{subfigure}[b]{0.77\columnwidth}
        \includegraphics[width=\linewidth]{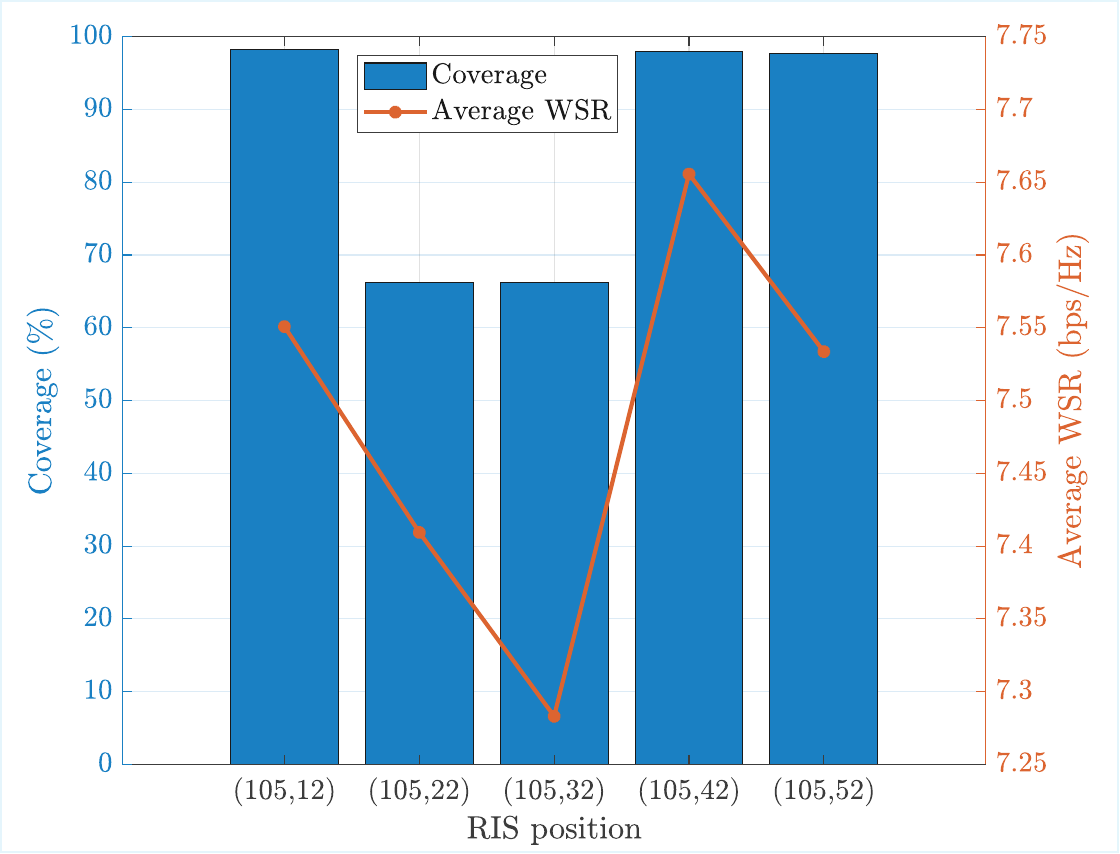}
        \caption{RIS positional analysis.}
        \label{fig:scenario4}
    \end{subfigure}    
    
    \begin{subfigure}[b]{0.75\columnwidth}
        \includegraphics[width=\linewidth]{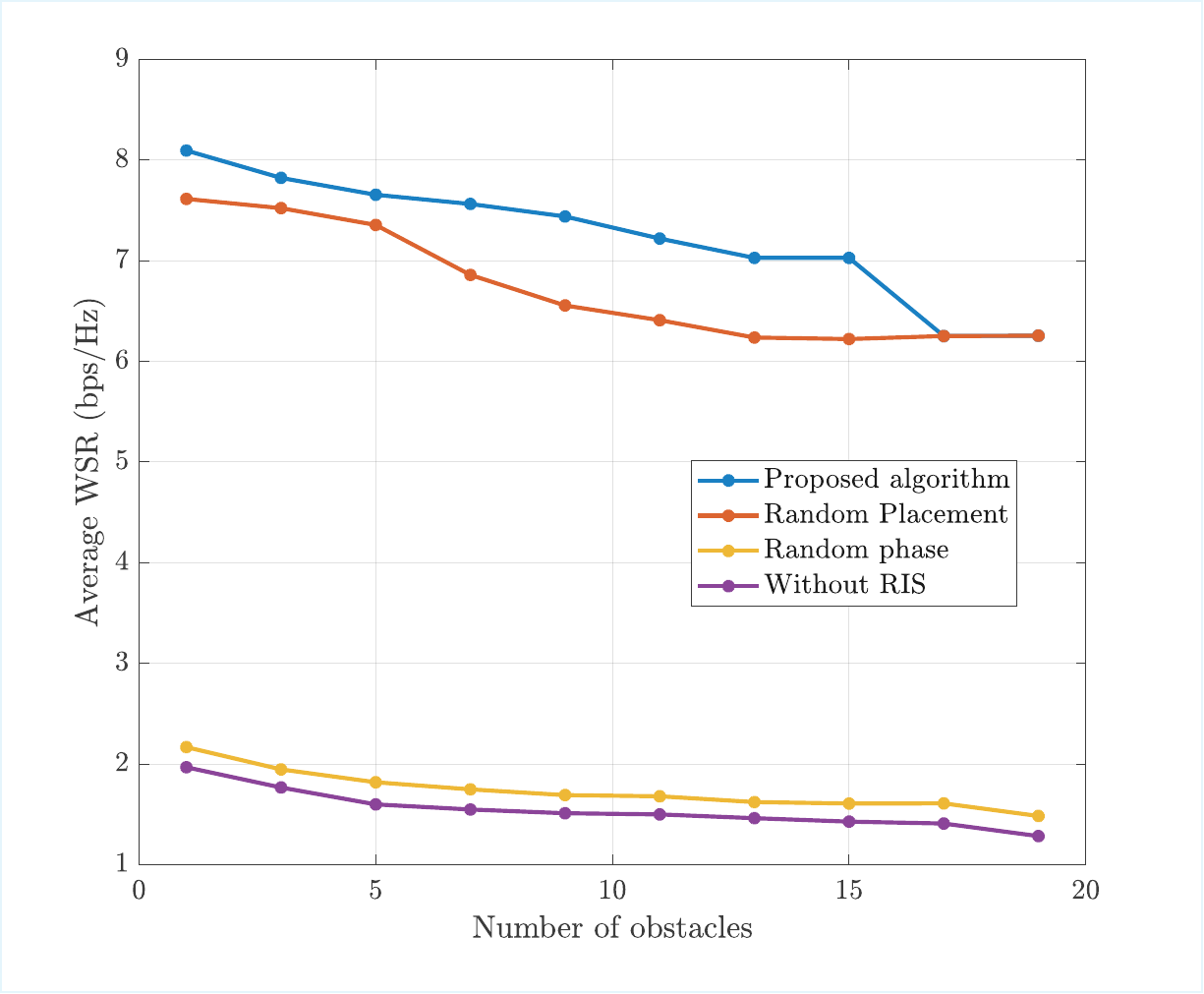}
        \caption{Obstacle analysis.}
        \label{fig:scenario4}
    \end{subfigure} 
    \caption{Further analysis of proposed algorithm for Scenario 2.}
\end{figure}

A major obstacle group is considered at the center of the cell in Scenario~2, as shown in Fig.~2~(b). A lower user density (on average, three fewer users) is assumed. The optimal location and average WSR are given in Table~\ref{tab:sim_params}. As shown in Table~\ref{tab:sim_params}, this extends coverage from 66.27\% to 97.91\% of the cell. Scenario~3, as shown in Fig.~2~(c), was designed to be difficult to extend coverage to the entire cell with a single RIS. The algorithm still derives a solution that covers regions shadowed by the major clusters and covers a portion of the region shadowed by the single wall obstacle (i.e., extends coverage to 97.21\% of the cell), with a similar WSR performance.


Finally, a practical Scenario~4 with four user hotspots at $C_1,C_2,C_3,C_4$ is considered as depicted in Fig.~2~(d). The scenario also assumes a spatially dependent $\lambda_u$, as given in Table~\ref{tab:sim_params}. Furthermore, although this scenario assumes the same obstacle configuration as in Scenario~3, the final solution shifts from (109~m, 45~m) to (114~m, 42~m) to accommodate the increase in user density in the region shadowed by the wall obstacle, highlighting the capability of the proposed method to converge to a location that not only maximizes coverage by covering most of the shadowed regions, but also effectively adapts to spatial variations in user density, resulting in improved WSR performance.





\subsection{Performance Analysis}

The average WSR vs.\ transmit power characteristics for Scenario~2 is shown in Fig.~4~(a). The performance of the proposed algorithm is much greater than the `without RIS' and the `random phase' cases, reaffirming the importance of an RIS in the system. In addition, the average WSR for deploying RIS in the optimal location  clearly exceeds the `Random Placement' simulation. This gain proves that optimal placement of RIS, not only results in an extension of coverage to about 98\% (or increase in fairness as per our objective $\mathbf{P1.1}$), but also results in an increase in the performance of the system. Furthermore, the WSR performance of the proposed scheme exceeds the performance of similar works \cite{imperfectcsiloc,indoorloc}. For instance, when $P_{max} = 40$ dBm, a gain of 3.4 and 2.7 bps/Hz is achieved over a multi-RIS assisted system that employs location optimization and hybrid beamforming for an indoor setup \cite{indoorloc} and a co-located multi-RIS assisted MISO system that employs joint location and beamforming optimization \cite{imperfectcsiloc}, respectively. It is also important to note that, owing to the optimality of the proposed scheme, 
this gain is achieved despite having a lower $M=16<32$, just a single RIS, practical number of users (average $K \approx 11 \gg2$), different obstacle configurations, and the randomness of users.

Fig.~4~(b) provides the convergence analysis of the inner optimization. For instance, the average WSR performance for $5$ AO iterations is obtained by averaging over multiple user instantiations, where the performance of each instantiation is evaluated by performing joint beamforming for 5 iterations. 

Then, Fig.~5~(a) provides a  detailed performance analysis of shifting the RIS across the vertical diameter of the cell in Scenario~2. As depicted, the converged solution (105~m, 42~m) maximizes both coverage and average WSR as desired.

Finally, Fig.~5~(b) generalizes Scenario~2 by varying the number of obstacles in the central cluster. As anticipated, performance degrades with the increase in number of obstacles. However, the rate of fall in performance of the proposed algorithm is much lesser than that of `random placement' due to the optimality of the proposed algorithm. Also, notably, when the number of obstacles in the central cluster is 17 or more (placed across the cell's vertical diameter), there is no way an RIS can help improve performance, forcing the proposed algorithm to perform like random placement.
\section{Conclusion}
\label{sec:Conclusion}
A novel method was proposed to tackle typical practical requirements of an RIS-assisted system, such as exploiting the user distribution to maximize performance, exploiting the obstacle configuration to maximize the coverage of the system, and handling user randomness. This is the first work, to the best of our knowledge, to handle the ideal hierarchical optimization directly - maximizing coverage, while also ensuring good performance. It is also notable that the method is valid for both perfect and imperfect CSI conditions depending on the adopted beamforming algorithm. A set of min-max solutions for the RIS placement was proposed to be obtained by searching over multiple discrete sets formed based on numerous realizations from the user distribution. Finally, a recursive strategy was used to find the final region to deploy the RIS. The superiority of the proposed algorithm was validated using simulations.

Notably, the proposed algorithm, both inner and outer optimizations, can be readily modified for a Multi-RIS assisted system. We look forward to investigating the same in our future work.

\end{document}